\documentclass[11pt]{article}
\usepackage{graphicx}
\usepackage{longtable,lscape}
\usepackage{amsfonts}
\usepackage{float}
\usepackage{url}

\newcommand{\captionfonts}{\footnotesize}
\makeatletter  
\long\def\@makecaption#1#2{%
  \vskip\abovecaptionskip
  \sbox\@tempboxa{{\captionfonts #1: #2}}%
  \ifdim \wd\@tempboxa >\hsize
    {\captionfonts #1: #2\par}
  \else
    \hbox to\hsize{\hfil\box\@tempboxa\hfil}%
  \fi
  \vskip\belowcaptionskip}
\makeatother 
\begin{document}
\title{A Quantum Cognition Analysis of the Ellsberg Paradox}
\author{Diederik Aerts, Bart D'Hooghe and Sandro Sozzo \vspace{0.5 cm} \\ 
        \normalsize\itshape
        Center Leo Apostel for Interdisciplinary Studies \\
        \normalsize\itshape
        Brussels Free University \\ 
        \normalsize\itshape
         Krijgskundestraat 33, 1160 Brussels, Belgium \\
        \normalsize
        E-Mails: \url{diraerts@vub.ac.be,bdhooghe@vub.ac.be} \\\url{ssozzo@vub.ac.be}
        }
\date{}
\maketitle              
\begin{abstract}
\noindent
The {\it expected utility hypothesis} is one of the foundations of classical approaches to economics and decision theory and Savage's {\it Sure-Thing Principle} is a fundamental element of it. It has been put forward that real-life situations exist, illustrated by the {\it Allais} and {\it Ellsberg paradoxes}, in which the Sure-Thing Principle is violated, and where also the expected utility hypothesis does not hold. We have recently presented strong arguments for the presence of a double layer structure, a {\it classical logical} and a {\it quantum conceptual}, in human thought and that the quantum conceptual mode is responsible of the above violation. We consider in this paper the Ellsberg paradox, perform an experiment with real test subjects on the situation considered by Ellsberg, and use the collected data to elaborate a model for the conceptual landscape surrounding the decision situation of the paradox. We show that it is the conceptual landscape which gives rise to a violation of the Sure-Thing Principle and leads to the paradoxical situation discovered by Ellsberg.  
\end{abstract}

\medskip
{\bf Keywords}: Sure-Thing Principle, Ellsberg paradox, conceptual landscape, quantum cognition

\vspace{-0.2 cm}

\section{Introduction\label{intro}}
In game theory, decision theory and economics the {\it expected utility hypothesis} requires that individuals evaluate uncertain prospects according to their expected level of `satisfaction' or `utility'. In particular, the expected utility hypothesis is the predominant descriptive and normative model of choice under uncertainty in economics. From a mathematical point of view the expected utility hypothesis is founded on the {\it von Neumann-Morgenstern utility theory} \cite{vonneumannmorgenstern1944}. These authors provided a set of `reasonable' axioms under which the expected utility hypothesis holds. One of the axioms proposed by von Neumann and Morgenstern is the {\it independence axiom} which is an expression of Savage's {\it Sure-Thing Principle} \cite{savage1954}, the latter being one of the building blocks of classical approaches to economics. Examples exist in the literature which show an inconsistency with the predictions of the expected utility hypothesis, namely a violation of the Sure-Thing Principle. These deviations, often called paradoxes, were firstly revealed by Maurice Allais \cite{allais1953} and Daniel Ellsberg \cite{ellsberg1961}. The Allais and Ellsberg paradoxes at first sight at least, indicate the existence of an {\it ambiguity aversion}, that is, individuals prefer `sure choices' over `choices that contain ambiguity'. Several attempts have been put forward to solve the drawbacks raised by the Allais and Ellsberg paradoxes but none of the arguments that have been proposed is, at the best of our knowledge, considered as conclusive.

The above problems are strongly connected with difficulties that afflict cognitive science, i.e. the concept combination problem (see, e.g., \cite{hampton1988}) and the disjunction effect (see, e.g., \cite{tverskyshafir1992}). It is indeed well known that concepts combine in human minds in such a way that they show deviations from the expectations that could be drawn in classical set and probability theories. Analogously, subjects take decisions which seem to contradict classical logic and probability theory. Trying to cope with these difficulties one of the authors has proposed, together with some co-workers, a formalism ({\it SCoP formalism}) in which context plays a relevant role in both concept combinations and decision processes \cite{gaboraaerts2002,aertsgabora2005a,aertsgabora2005b}. Moreover, this role is very similar to the role played by the (measurement) context on microscopic particles in quantum mechanics. Within the SCoP perspective models have been elaborated which use the mathematical formalism of quantum mechancis to describe both the concept combinations and the disjunction effect, and which accord with the experimental data existing in the literature \cite{aerts2009,aertsdhooghe2009,aerts2007,aerts2010}. This analysis has allowed the authors to suggest the hypothesis that two structured and superposed layers can be identified in human thought: a {\it classical logical layer}, that can be modeled by using a classical Kolmogorovian probablity framework, and a {\it quantum conceptual layer}, that can instead be modeled by using the probabilistic formalism of quantum mechanics. The thought process in the latter layer is given form under the influence of the totality of the surrounding conceptual landscape, hence context effects are fundamental in this layer. The relevance of the quantum conceptual layer in producing the disjunction effect will be extensively discussed in a forthcoming paper \cite{aertsczachordhooghe2011}. In the present paper we instead focus on the Ellsberg paradox. More precisely, after introducing Savage's Sure-Thing Principle and its violation occurring in the Ellsberg paradox in Sec. \ref{ellsberg}, we provide in Sec. \ref{preliminary} a preliminary analysis of the paradox, clarifying and fixing, in particular, some assumptions that are not made explicit in the standard presentations of it. Then, we discuss in Sec. \ref{experiment} a real experiment on 59 test subjects that we have performed to test the Ellsberg paradox, and examine the obtained results. More specifically, we identify from the obtained answers and explanations the conceptual landscapes that we consider relevant in formulating the paradox. We finally work out in Sec. \ref{quantum} the structural mathematical scheme for a quantum model in which each conceptual landscape is represented by a vector of a Hilbert space and the qualitative results obtained in our experiment are recovered by considering the overall conceptual landscape as the superposition of the single landscapes.   

We conclude this section with two remarks. Firstly, we note that in our approach the explanation of the violation of the expected utility hypothesis and the Sure-Thing Principle is not (only) the presence of an ambiguity aversion. On the contrary, we argue that the above violation is due to the concurrence of superposed conceptual landscapes in human minds, of which some might be linked to ambiguity aversion, but other completely not. We therefore maintain that the violation of the Sure-Thing Principle should not be considered as a fallacy of human thought, as often claimed in the literature but, rather, as the proof that real subjects follow a different way of thinking than the one dictated by classical logic in some specific situations, which is context-dependent. Secondly, we observe that an explanation of the violation of the expected utility hypothesis and the Sure-Thing Principle in terms of quantum probability and quantum interference has already been presented in the literature (see, e.g., \cite{busemeyerwangtownsend06,franco07,khrennikovhaven09,pothosbusemeyer09}). What is new in our approach is the fact that the quantum mechanical modeling is not just an elegant formal tool but, rather, it reveals the presence of an undelying quantum conceptual thought.

\vspace{-0.2 cm}  

\section{The Sure-Thing Principle and the Ellsberg Paradox\label{ellsberg}}
Savage introduced the {\it Sure-Thing Principle} \cite{savage1954} inspired by the following story.

{\it A businessman contemplates buying a certain piece of property. He considers the outcome of the next presidential election relevant. So, to clarify the matter to himself, he asks whether he would buy if he knew that the Democratic candidate were going to win, and decides that he would. Similarly, he considers whether he would buy if he knew that the Republican candidate were going to win, and again finds that he would. Seeing that he would buy in either event, he decides that he should buy, even though he does not know which event obtains, or will obtain, as we would ordinarily say.} 

The Sure-Thing Principle is equivalent to the independence axiom of expected utility theory: `independence' here means that if persons are indifferent in choosing between simple lotteries $L_{1}$ and $L_{2}$, they will also be indifferent in choosing between $L_{1}$ mixed with an arbitrary simple lottery $L_{3}$ with probability $p$ and $L_{2}$ mixed with $L_{3}$ with the same probability $p$.

Let us consider the situation put forward by Daniel Ellsberg \cite{ellsberg1961} to point out an inconsistency with the predictions of the expected utility hypothesis and a violation of the Sure-Thing Principle. Imagine an urn known to contain 30 red balls and 60 balls that are either black or yellow, the latter in unknown proportion. One ball is to be drawn at random from the urn. To `bet on red' means that you will receive a prize $a$ (say, 10 euros) if you draw a red ball (`if red occurs') and a smaller amount $b$ (say, 0 euros) if you do not. If test subjects are given the following 4 options: (I) `a bet on red', (II) `a bet on black', (III) `a bet on red or yellow', (IV) `a bet on black or yellow', and are then presented with the choice between bet I and bet II, and the choice between bet III and bet IV, it appears that a very frequent pattern of response is that bet I is preferred to bet II, and bet IV is preferred to bet III. This violates the Sure-Thing Principle, which requires the ordering of I to II to be preserved in III and IV (since these two pairs differ only in the pay-off when a yellow ball is drawn, which is constant for each pair). The first pattern, for example, implies that test subjects bet on red rather than on black; and also that they will bet against red rather than against black.

The contradiction above suggests that preferences of `real-life' subjects are inconsistent with Savage's Sure-Thing Principle of expected utility theory. A possible explanation of this drawback could be that people make a mistake in their choice and that the paradox is caused by an error of reasoning. In our view, however, these examples show that subjects make their decisions in ways which do violate the Sure-Thing Principle, but not because they make an error of reasoning, but rather because they follow a different type of reasoning which is not only guided by logic but also by conceptual thinking which is structurally related to quantum mechanics. We stress that in the Ellsberg paradox the situation where the number of yellow balls and the number of black balls are not known individually, only their sum being known to be 60, introduces the so-called {\it disjunction effect} \cite{tverskyshafir1992}, which will be systematically discussed by us in a forthcoming paper \cite{aertsczachordhooghe2011}.

\vspace{-0.2 cm}

\section{A preliminary analysis of the paradox\label{preliminary}}
Frank Knight introduced the distinction between different types of uncertainty \cite{knight1921}, and Daniel Ellsberg stimulated the reflection about these different types of uncertainty \cite{ellsberg1961}. More explicitly, Ellsberg puts forward the notion of {\it ambiguity} as an uncertainty without any well-defined probability measure to model this uncertainty, as opposed to {\it risk}, where such a probability measure does exist. In the case of the Ellsberg paradox situation, `betting on red' concerns a situation in which the uncertainty is modeled by a probability measure which is given, namely a probability of $\frac{1}{3}$ to win the bet, and a probability of $\frac{2}{3}$ to lose it. For `betting on black', however, the situation is such that no definite probability measure models the situation related to this bet. Indeed, since it is only known that the sum of the black and the yellow balls is 60, the number of black balls is not known. If no additional information is given specifying in more detail the situation of the Ellsberg paradox, `betting on black' will be a situation of ambiguity, since the probability measure associated with this bet is not known. Of course, by making a specific additional assumption, namely the assumption that black and yellow balls are chosen at random until their sum reaches 60, we can re-introduce a probability measure corresponding to the `bet on black' situation. In this case, also for `betting on black' the probability of winning equals $\frac{1}{3}$ and that of losing equals $\frac{2}{3}$. If the Ellsberg paradox situation is presented as a real-life situation, for reasons of symmetry, it can be supposed that indeed black and yellow balls are chosen at random until their sum reaches 60, and then put in the urn. In this case a `bet on black' is equivalent with a `bet on red'. 

However, there are many possible situations of `real life' where this symmetry is perhaps not present, one obvious example being the one where the person proposing to bet following an Ellsberg type of situation has the intention to trick, and for example installs a way to have systematically less black balls than yellow balls in the urn. Of course, the real aim of the Ellsberg paradox is to show that `people will already take into account this possibility' even if nothing is mentioned extra, which means that most probably the situation is symmetric. We will see that our analysis by means of the introduction of different conceptual landscapes sheds light on this aspect of the paradox.

In the following we put forward an analysis of the Ellsberg paradox situation, using the explanation we introduced for the presence of underextension and overextension for concept combinations and for the disjunction effect \cite{aerts2009,aertsdhooghe2009}. The essential element of our explanation is the distinction between `the conceptual landscape surrounding a given situation' and the `physical reality of this given situation'. The probabilities governing human decisions are related to this conceptual landscape and not necessarily to the physical reality of a given situation. Although there is a correspondence between the physical reality of a situation and the surrounding conceptual landscape, in most cases this correspondence is far from being an isomorphism. For the situation of the Ellsberg paradox, let us first describe the physical reality of the situation and then provide a plausible conceptual landscape surrounding this situation.

The physical situation is the urn containing red, black and yellow balls, with the number of red balls being 30 and the sum of the number of black balls and yellow balls being 60. The original article \cite{ellsberg1961} does not specify the physical situation in any further detail, leaving open the question as to `how the black and the yellow balls are chosen when 60 of them are put in the urn'. We prefer to make the physical situation more specific and introduce an additional hypothesis, namely that the black and the yellow balls are put in the urn according to a coin toss. When heads is up, a black ball is added to the urn, and when tails is up a yellow ball is added. 

Prepared according to the Ellsberg situation, the urn will contain 30 red balls, $60-n$ black balls and $n$ yellow balls, where $n \in \{0, 1, \ldots, 59, 60\}$. In this case, when we choose a ball at random, there is a probability of $\frac{1}{3}$ for a red ball to turn up, a probability of $\frac{60-n}{90}$ for a black ball to turn up, and a probability of $\frac{n}{90}$ for a yellow ball to turn up. For an urn prepared according to the outcome of a coin toss, however, the probability for red to turn up is $\frac{1}{3}$, the probability for black to turn up is $\frac{1}{3}$, and the probability for yellow to turn up is also $\frac{1}{3}$.

\vspace{-0.2 cm}

\section{An experiment testing the Ellsberg paradox\label{experiment}}
For the type of analysis we make, we need to account for different pieces of conceptual landscape. To gather relevant information, we decided to perform a test of the Ellsberg paradox problem. To this end, we sent out the following text to several friends, relatives and students. We asked them to forward our request to others, so that our list could also include people we didn't know personally.

\vspace{.2cm}
{\it We are conducting a small-scale statistics investigation into a particular problem and would like to invite you to participate as test subjects. Please note that it is not the aim for this problem to be resolved in terms of correct or incorrect answers. It is your preference for a particular choice we want to test. The question concerns the following situation. 

Imagine an urn containing 90 balls of three different colors: red balls, black balls and yellow balls. We know that the number of red balls is 30 and that the sum of the the black balls and the yellow balls is 60. The questions of our investigation are about the situation where somebody randomly takes one ball from the urn. 

- The first question is about a choice to be made between two bets: bet I and bet II. Bet I involves winning `10 euros when the ball is red' and `zero euros when it is black or yellow'. Bet II involves winning `10 euros when the ball is black' and `zero euros when it is red or yellow'. The first question we would ask you to answer is: Which of the two bets, bet I or bet II, would you prefer? 

- The second question is again about a choice between two different bets,  bet III and bet IV. Bet III involves winning `10 euros when the ball  is red or yellow' and `zero euros when the ball is black'.  Bet IV involves winning `10 euros when the ball is black or  yellow' and `zero euros when the ball is red'. The  second question therefore is: Which of the two bets, bet III or bet IV, would you  prefer?

Please provide in your reply message the following information: 

For question 1, your preference (your choice between bet I and bet II). For question 2, your preference (your choice between bet III and bet IV). 

By `preference' we mean `the bet you would take if this situation happened to you in real life'. You are expected to choose one of the bets for each of the questions, i.e. `not choosing is no option'. 

You are welcome to provide a brief explanation of your preferences, which may be of a purely intuitive nature, only mentioning feelings, for example, but this is not required. It is allright if you only state your preferences without giving any explanation.

One final remark about the colors. Your choices should not be affected by any personal color preference. If you feel that the colors of the example somehow have an influence on your choices, you should restate the problem and take colors that are indifferent to you or, if this does not work, other neutral characteristics to distinguish the balls.} 

\vspace{.2cm}
Let us now analyze the obtained results.

We had 59 respondents participating in our test of the Ellsberg paradox problem, of whom 34 preferred bets $I$ and $IV$, 12 preferred bets $II$ and $III$, 7 preferred bets $II$ and $IV$ and 6 preferred bets $I$ and $III$. This makes the weights with preference of bet $I$ over bet $II$ to be 0.68 against 0.32, and the weights with preference of bet $IV$ over bet $III$ to be 0.71 against 0.29. It is interesting to note that 34+12=46 people chose the combination of bet $I$ and bet $IV$ or bet $II$ and bet $III$, which is $87\%$. Of the 59 participants there were 10 who provided us an explanation for their choice. Interestingly, an independent consideration of this group of 10 reveals a substantial deviation of their statistics from the overall statistics: only 4 of them chose bet $I$ and bet $IV$, 2 chose bet $II$ and bet $III$, 3 chose bet $II$ and bet $IV$, and 1 chose bet $I$ and bet $III$. What is even more interesting, however, is that only half of them preferred bet $I$ to bet $II$. So the participants in the `explaining sub-group' were as likely to choose bet $I$ as they were likely to choose bet $II$. This is too small a sample of `subjects providing an explanation' to be able to make a firm conclusion about the different pieces of conceptual landscape in this Ellsberg paradox situation. Since this article is mainly intended to illustrate our way of modeling the situation, we will make a proposal for such a possible conceptual landscape.

A first piece of conceptual landscape is: `an urn is filled with 30 balls that are red, and with 60 balls chosen at random from a collection of black and a collection of yellow balls'. We call this piece of conceptual landscape the {\it Physical Landscape}. It represents that which is most likely to correspond to the physical presence of an actual Ellsberg paradox situation. A second piece of conceptual landscape is: `there might well be substantially fewer black balls than yellow balls in the urn, and so also substantially fewer black balls than red balls'. We call this piece of conceptual landscape the {\it First Choice Pessimistic Landscape}. It represents a guess of a less advantage situation compared to the neutral physical one, when the subject is reflecting on the first choice to be made. A third piece of conceptual landscape is: `there might well be substantially more black balls than yellow balls in the urn, and so also substantially more black balls than red balls'. This third piece we call the {\it First Choice Optimistic Landscape}. It represents a guess of a more advantage situation compared to the neutral physical one, when the subject is reflecting on the first choice to be made. A fourth piece of conceptual landscape is: `there might well be substantially fewer yellow balls than black balls, and so substantially fewer red plus yellow balls than black plus yellow balls, of which there are a fixed number, namely 60'. This fourth piece we call {\it Second Choice Pessimistic Landscape}. It represents a guess of a less advantage situation compared to the neutral physical one, when the subject is reflecting on the second choice to be made.  A fifth piece of conceptual landscape is: `there might well be substantially more yellow balls than black balls, and so substantially more red plus yellow balls than black plus yellow balls, of which there are a fixed number, namely 60'. This fifth piece we call the {\it Second Choice Optimistic Landscape}. It represents a guess of a more advantage situation compared to the neutral physical one, when the subject is reflecting on the second choice to be made. A sixth piece of conceptual landscape, which we call the {\it Suspicion Landscape}, is: `who knows how well the urns has been prepared, because after all, to put in 30 red balls is straightforward enough, but to pick 60 black and yellow balls is quite another thing; who knows whether this is a fair selection or somehow a biased operation, there may even have been some kind of trickery involved'. A seventh piece of conceptual landscape is: `if things become too complicated I'll bet on the simple situation, the situation I understand well', which we call the {\it Don't Bother Me With Complications Landscape}.

These pieces of conceptual landscape are the ones we can reconstruct taking into account the explanations we received from our test subjects. We are convinced, however, that they are by no means the only possible relevant pieces of conceptual landscape. For example, one of the subjects who participated in our test and chose bet II and bet III said that she would have chosen differently, preferring bet I and bet IV, if more money had been involved. This leads us to believe that what plays a major role too in the choices the subjects make is whether they regard the test as a kind of funny game or make a genuine attempt to try and guess what they would do in real life when presented with a betting situation of the Ellsberg type. At an even more subtle level, subjects who feel that by choosing the combination bet I and bet IV, they would be choosing for a greater degree of predictability, might be tempted to change their choice, preferring the more unpredictable combination of bet II and bet III, because this is intellectually more challenging, although again this would depend on how they conceive the situation. Indeed, we firmly believe that the determining of further conceptual landscapes that are relevant involves even more subtle aspects. 

\vspace{-0.2 cm}

\section{A quantum model for conceptual landscapes\label{quantum}}
Using the conceptual landscapes we did mention explicitly, let us illustrate in this section how a quantum modeling scheme can be worked out.
Consider the piece of conceptual landscape which we called the {\it Physical Landscape}, and suppose that it is the only piece, i.e. that it constitutes the whole conceptual landscape for a specific individual subject. This means this subject has no preference for bet I or bet II, and also has no preference for bet III or bet IV, so that the Sure-Thing Principle is not violated. A simple quantum mechanical model of this situation is one where we represent the conceptual landscape by means of vector $|A\rangle$, and the choice between bet I and bet II by means of a projection operator $M$, such that 
\begin{equation}
\mu_M(A)=\langle A|M|A\rangle
\end{equation}
is the weight for a subject to choose bet I, while $1-\mu_M(A)=\langle A|1-M|A\rangle$ is the weight for a subject to choose bet II, while the choice between bet III and bet IV is described by a projection operator $N$, such that 
\begin{equation}
\mu_N(A)=\langle A|N|A\rangle
\end{equation}
is the weight for a subject to choose bet III, while $1-\mu_N(A)=\langle A|1-N|A\rangle$ is the weight for a subject to choose bet IV. We have
\begin{equation}
\mu_M(A)=\mu_N(A)=\frac{1}{2}.
\end{equation}
Consider now the piece of conceptual landscape {\it First Choice Pessimistic}, and suppose that this is the only piece of conceptual landscape. Then bet I will be strongly preferred over bet II, and a quantum modeling of this situation consists in representing this piece of conceptual landscape by means of a vector $|B\rangle$ such that $\mu_M(B)=\langle B|M|B\rangle$ and $1-\mu_M(B)=\langle B|1-M|B\rangle$ represent the weights for subjects to choose bet I and bet II, respectively, so that $1-\mu_M(B) \ll \mu_M(B)$ or, equivalently, $\frac{1}{2} \ll \mu_M(B)$. It is not easy to know how $\mu_N(B)$ will be under conceptual landscape {\it First Choice Pessimistic}. Indeed, our experience with the test we conducted indicates that when subjects are asked to compare bet III and bet IV, other conceptual landscapes become relevant and predominant than the conceptual landscapes that are relevant and predominant when they are asked to compare bet I and bet II. Subjects who tend to give a high weight to conceptual landscape {\it First Choice Pessimistic} when comparing bet I and bet II, i.e. `who fear that there might be substantially fewer black balls than red balls' seem to focus rather on the variability of the yellow balls when asked to compare bet III and bet IV, and tend to give dominance to conceptual landscape {\it Second Choice Pessimistic}, `fearing that there might be substantially fewer yellow balls than black balls, and hence also fewer red plus yellow balls than black plus yellow balls'. This is borne out by the fact that 46 people, or $87\%$ of the total number of participants, choose for the combination of bet $I$ and bet $IV$ or bet $II$ and bet $III$. However, we also noted that some subjects gave dominance to what we have called conceptual landscape {\it Don't Bother Me With Complications} when they were asked to choose between bet III and bet IV. They had preferred bet I to bet II, and now also preferred bet III to bet IV. When asked why they preferred bet I to bet II, their answer was `because we know what the risk is for red, but for black we do not'. Interestingly, when we asked them to reconsider their choice with respect to bet III and bet IV -- they had preferred bet III -- now explaining to them that bet IV gave rise to `less uncertainty' than bet III, they remained with their preference for bet III to bet IV, commenting that `anyhow betting on red made them feel more comfortable much like when asked to choose between bet I and bet II'. We believe that the rather artificial aspect of choosing between bet III and bet IV, of considering outcomes whose definitions are disjunctions of simple outcomes, makes this choice essentially more complicated, such that the choices made by these subjects are in line with what the Ellsberg paradox analysis tries to put forward. However, due to the relatively greater complexity of bet III and bet IV, as compared to bet I and bet II, this aspect is not revealed.

Anyhow, considerations like the one above are not our primary concern here, since we mainly want to give an account of how we apply our quantum-conceptual modeling scheme in the situation we have described. Again, because of the rather limited nature of the experiment conducted for this article, we have not been able to estimate the value of $\mu_N(B)$. However, if we call $|D\rangle$ the vector representing the conceptual landscape {\it Second Choice Pessimistic}, we have $\mu_N(D) \ll \frac{1}{2}$. If $|C\rangle$ and $|E\rangle$ represent the {\it First Choice Optimistic Landscape} and the {\it Second Choice Optimistic Landscape}, we have $\mu_M(C) \ll \frac{1}{2}$ and $\frac{1}{2} \ll \mu_N(E)$, respectively.

Now if we look at the {\it Suspicion Landscape} and the  {\it DonÕt Bother Me With Complications Landscape} -- let us represent them by vectors $|F\rangle$ and $|G\rangle$, respectively -- we have a situation that resembles much more what is generally claimed with respect to the Ellsberg paradox situation, namely $\mu_M(F) \ll \frac{1}{2}$ and $\frac{1}{2} \ll \mu_N(F)$, and also $\mu_M(G) \ll \frac{1}{2}$ and $\frac{1}{2} \ll \mu_N(G)$, which both induce a direct violation of the Sure-Thing Principle. Following the general quantum modeling scheme we worked out in detail in earlier publications \cite{gaboraaerts2002,aertsgabora2005a,aertsgabora2005b,aerts2009,aertsdhooghe2009,aerts2007,aerts2010}, when all these pieces of conceptual landscape are present with different weights, the vector to model this situation is a normalized superposition of the vectors $|A\rangle, |B\rangle, |C\rangle, |D\rangle, |E\rangle, |F\rangle$ and $|G\rangle$. This makes it possible to choose coefficients of superposition such that if the Ellsberg paradox situation is surrounded by the conjunction of all these pieces of conceptual landscape, the Sure-Thing Principle will be violated in a way corresponding to experimental data that are collected with respect to this situation.

At this stage we should point out the following. In our earlier work \cite{aerts2009,aertsdhooghe2009} we introduced the notion of `conceptual landscape'. One of the reasons for this is that our approach is grounded in a modeling of concepts and their combinations \cite{gaboraaerts2002,aertsgabora2005a,aertsgabora2005b}. Of course, a human decision will be sometimes made under the influence of a surrounding that cannot easily be expressed as a conceptual landscape. When we pointed out explicitly that it is `the whole' conceptual landscape which needs to be taken into account and modeled within our quantum modeling scheme, we could have expressed this in a more complete way by using the notion of `worldview' as understood and analyzed in detail in the Leo Apostel Center \cite{aertsaposteldemoorhellemansmaexvanbellevanderveken1994,aertsaposteldemoorhellemansmaexvanbellevanderveken1995,aertsvanbellevanderveken1999}. These are all possible elements of a worldview surrounding a given situation that can influence a human decision being made in this situation. Of course, if the elements of the surrounding worldview considered can be expressed conceptually, i.e. if they are of the form of a conceptual landscape, then these elements can be taken into account by means of the quantum modeling scheme we have developed in earlier work for concepts and their combinations \cite{gaboraaerts2002,aertsgabora2005a,aertsgabora2005b,aerts2009,aertsdhooghe2009,aerts2007,aerts2010}. This being the case, we are already able to grasp an enormously important and also substantial part of the dynamics generated by the totality of the worldview influence, which is the reason why we have focused on this in our previous work.

Let us close this paper with a final remark. We believe that the attempts of economists and psychologists to distinguish between risk and ambiguity \cite{einhornhogarth1978,einhornhogerth1985,budescuwallsten1987,kahnsarin1988,heathtversky1991,schoemaker1990,winkler1991,bromileycurley1992,camererweber1992,ghoshray1992,ghoshcrain1993}, also by referring to situations where a probability measure exists (risk) or does not exist (ambiguity) \cite{ellsberg1961,knight1921,kahnsarin1988,hogarth1989,fishburn1990,frischbaron1988,lopes1983}, constitute a very interesting approach that deserves further development. Indeed, recent research  has demonstrated that different parts of a subject's brain are activated depending on whether they are confronted with a situation of risk or a situation of ambiguity \cite{huettelstowegordonwarnerplatt2006}. In this respect, we recall that we have elaborated an axiomatic approach to quantum probability called `the hidden measurement approach', considering situations with existing probability measures, but such that the probability measures depend on the measurement situation, which makes them contextual \cite{aerts1986,aerts1993,aerts1995,aertsaerts1997,aertsaertscoeckedhooghedurtvalckenborgh1997,aerts1998}. What economists have considered a mathematical approach to ambiguity \cite{ellsberg1961,knight1921,kahnsarin1988,hogarth1989,fishburn1990,frischbaron1988,lopes1983} bears a great similarity to aspects of this hidden measurement approach \cite{aerts1986,aerts1993,aerts1995,aertsaerts1997,aertsaertscoeckedhooghedurtvalckenborgh1997,aerts1998}, and a further formal development of a probability theory compatible with this idea of contextuality has led us to introduce a generalized probability formalism \cite{aerts2002}. In future work we intend to investigate how this allows to cope with specific types of ambiguity in a mathematically rigorous way.

\medskip
\noindent
{\bf Acknowledgments.} The authors are greatly indebted with the 59 friends and colleagues for participating in the experiment. This research was supported by Grants G.0405.08 and G.0234.08 of the Flemish Fund for Scientific Research.


\vspace{-0.2 cm}

\begin{thebibliography}{99} 

\bibitem{vonneumannmorgenstern1944} von Neumann, J., Morgenstern, O.: Theory of Games and Economic Behavior. Princeton University Press, Princeton (1944)

\bibitem{savage1954} Savage, L.J.: The Foundations of Statistics. Wiley, New York (1954)

\bibitem{allais1953} Allais, M.: Le Comportement de l'Homme Rationnel Devant le Risque: Critique des Postulats et Axiomes de l'\'{E}cole Am\'{e}ricaine. Econometrica, 21, 503--546 (1953)

\bibitem{ellsberg1961} Ellsberg, D.: Risk, Ambiguity, and the Savage Axioms. Quart. J. Econ., 75(4), 643--669 (1961)

\bibitem{hampton1988} Hampton, J.A.: Disjunction of Natural Concepts. Memory \& Cognition, 16, 579--591 (1988)

\bibitem{tverskyshafir1992} Tversky, A., Shafir, E.: The Disjunction Effect in Choice Under Uncertainty. Psych. Sci., 3, 305--309 (1992)

\bibitem{gaboraaerts2002} Gabora, L., Aerts, D.: Contextualizing Concepts Using a Mathematical Generalization of the Quantum Formalism. J. Exp. Theor. Art. Int., 14, 327--358 (2002)

\bibitem{aertsgabora2005a} Aerts, D., Gabora, L.: A Theory of Concepts and Their Combinations I: The Structure of the Sets of Contexts and Properties. Kybernetes, 34, 167--191 (2005)

\bibitem{aertsgabora2005b} Aerts, D., Gabora, L.: A Theory of Concepts and Their Combinations II: A Hilbert Space Representation. Kybernetes, 34, 192--221 (2005)

\bibitem{aerts2009} Aerts, D.: Quantum Structure in Cognition. J. Math. Psych., 53, 314--348 (2009)

\bibitem{aertsdhooghe2009} Aerts, D., D'Hooghe, B.: Classical Logical Versus Quantum Conceptual Thought: Examples in Economics, Decision Theory and Concept Theory. In: Bruza, P.D., Sofge, D., Lawless, W., van Rijsbergen, C.J., Klusch, M.  (eds.) Proceedings of QI 2009-Third International Symposium on Quantum Interaction. LNCS, vol. 5494, pp. 128--142. Springer, Berlin, Heidelberg (2009)

\bibitem{aerts2007} Aerts, D.: Quantum Interference and Superposition in Cognition: Development of a Theory for the Disjunction of Concepts. Archive reference and link: \url{http://arxiv.org/abs/0705.1740} (2007). In: Aerts, D. D'Hooghe, B., Note, N. (eds.) Worldviews, Science and Us: Bridging Knowledge and Its Implications for Our Perspectives of the World, in print. World Scientific, Singapore (2011) 

\bibitem{aerts2010} Aerts, D.: General Quantum Modeling of Combining Concepts: A Quantum Field Model in Fock Space.  Archive reference and link: \url{http://arxiv.org/abs/0705.1740} (2007)


\bibitem{aertsczachordhooghe2011} Aerts, D., Broekaert, J., Czachor, M., D'Hooghe, B.: The Violation of Expected Utility Hypothesis in the Disjunction Effect. Submitted to Proceedings of QI 2011-Fifth International Symposium on Quantum Interaction (2011)





\bibitem{busemeyerwangtownsend06} Busemeyer, J.R., Wang, Z., Townsend J.T.: Quantum Dynamics of Human Decision-Making. J. Math. Psych., 50, 220--241 (2006)

\bibitem{franco07} Franco, R.: Risk, Ambiguity and Quantum Decision Theory. Archive reference and link: \url{http://arxiv.org/abs/0711.0886} (2007)

\bibitem{khrennikovhaven09} Khrennikov, A.Y., Haven, E.: Quantum Mechanics and Violations of the Sure-Thing Principle: The Use of Probability Interference and Other Concepts. J. Math. Psych., 53, 378--388 (2009)

\bibitem{pothosbusemeyer09} Pothos, E.M., Busemeyer, J.R.: A Quantum Probability Explanation for Violations of `Rational' Decision Theory. Proc. Roy. Soc. B, 276, 2171--2178 (2009)

\bibitem{knight1921} Knight, F.H.: Risk, Uncertainty and Profit. Houghton Mifflin, Boston (1921)

\bibitem{aertsaposteldemoorhellemansmaexvanbellevanderveken1994} Aerts, D., Apostel, L., De Moor, B., Hellemans, S., Maex, E., Van Belle, H., Van der Veken, J.: Worldviews, from Fragmentation towards Integration. VUB Press, Brussels (1994)

\bibitem{aertsaposteldemoorhellemansmaexvanbellevanderveken1995} Aerts, D., Apostel, L., De Moor, B., Hellemans, S., Maex, E., Van Belle, H., Van der Veken, J.: Perspectives on the World, an Interdisciplinary Reflection. VUB Press, Brussels (1995)


\bibitem{aertsvanbellevanderveken1999} Aerts, D., Van Belle, H., Van der Veken, J. (eds.) Worldviews and the Problem of Synthesis. Springer, Dordrecht (1999)

\bibitem{einhornhogarth1978} Einhorn, H.J., Hogarth, R.M.: Confidence in Judgment: Persistence of the Illusion of Validity. Psych. Rev., 85, 395--416 (1978)

\bibitem{einhornhogerth1985} Einhorn, H.J., Hogarth, R.M.: Ambiguity and Uncertainty in Probabilistic Inference. Psych. Rev., 92, 433--461 (1985)

\bibitem{budescuwallsten1987} Budescu, D.V., Wallsten, T.S.: Subjective Estimation of Precise and Vague Uncertainties. In: Wright, G., Ayton, P. (eds.) Judgmental Forecasting. Wiley, New York (1987)

\bibitem{kahnsarin1988} Kahn, B.E., Sarin, R.K.: Modeling Ambiguity in Decisions Under Uncertainty. J. Cons. Res., 15, 265--272 (1988)

\bibitem{heathtversky1991} Heath, C., Tversky, A.: Preference and Belief: Ambiguity and Competence in Choice Under Uncertainty. J. Risk Unc., 4, 5--28 (1991)

\bibitem{schoemaker1990} Schoemaker, P.: Are Risk Attitudes Related Across Domains and Response Modes?. Man. Sci., 36, 1451--1463 (1990)

\bibitem{winkler1991} Winkler, R.: Ambiguity, Probability, Preference, and Decision Analysis. J. Risk Unc., 4, 285--297 (1991)

\bibitem{bromileycurley1992} Bromiley, P., Curley, S.: Individual Differences in Risk Taking. In: Yates, J. (ed.) Risk Taking Behavior. pp. 87--132. Wiley, New York (1992)

\bibitem{camererweber1992} Camerer, C., Weber, M.: Recent Developments in Modeling Preferences: Uncertainty and Ambiguity. J. Risk Unc., 5, 325--370 (1992)

\bibitem{ghoshray1992} Ghosh, D., Ray, M.R.: Risk Attitude, Ambiguity, Intolerance and Decision Making: An Exploratory Investigation. Dec. Sci., 23, 431--444 (1992)

\bibitem{ghoshcrain1993} Ghosh, D., Crain, T.L.: Structure of Uncertainty and Decision Making: An Experimental Investigation. Dec. Sci., 24, 789--807 (1993) 

\bibitem{hogarth1989} Hogarth, R.M.: Ambiguity in Competitive Decision Making: Some Implications and Tests. In: Fishburn, P.C., Lavelle, I. (eds.) Annals of Operations Research. vol. 19, pp. 31--50 (1989)

\bibitem{fishburn1990} Fishburn, P.C.: On the Theory of Ambiguity. Working Paper, AT\&T Bell Laboratories, NJ (1990)

\bibitem{frischbaron1988} Frisch, D., Baron, J.: Ambiguity and Rationality. J. Bus. Dec. Mak., 1, 149--157 (1988)

\bibitem{lopes1983} Lopes, L.L.: Some Thoughts on the Psychological Concept of Risk. J. Exp. Psych., 9, 137--144 (1983)

\bibitem{huettelstowegordonwarnerplatt2006} Huettel, S.A., Stowe, C.J., Gordon, E.M., Warner, B.T., Platt, M.L.: Neural Signatures of Economic Preferences for Risk and Ambiguity. Neuron, 49, 765--775 (2006)

\bibitem{aerts1986} Aerts, D.: A Possible Explanation for the Probabilities of Quantum Mechanics. J. Math. Phys., 27, 202--210 (1986)

\bibitem{aerts1993} Aerts, D.: Quantum Structures Due to Fluctuations of the Measurement Situations. Int. J. Theor. Phys., 32, 2207--2220 (1993)

\bibitem{aerts1995} Aerts, D.: Quantum Structures: An Attempt to Explain Their Appearance in Nature. Int. J. Theor. Phys., 34, 1165--1186 (1995)

\bibitem{aertsaerts1997} Aerts, D. and Aerts, S.: The Hidden Measurement Formalism: Quantum Mechanics as a Consequence of Fluctuations on the Measurement. In: Ferrero, M., van der Merwe, A. (eds.) New Developments on Fundamental Problems in Quantum Physics. pp. 1--6. Springer, Dordrecht (1997) 

\bibitem{aertsaertscoeckedhooghedurtvalckenborgh1997} Aerts, D., Aerts, S., Coecke, B., D'Hooghe, B., Durt, T., Valckenborgh, F.: A Model with Varying Fluctuations in the Measurement Context. In: Ferrero, M., van der Merwe, A. (eds.) New Developments on Fundamental Problems in Quantum Physics. pp. 7--9. Springer, Dordrecht (1997)

\bibitem{aerts1998} Aerts, D.: The Hidden Measurement Formalism: What Can Be Explained and Where Paradoxes Remain. Int. J. Theor. Phys., 37, 291--304 (1998)

\bibitem{aerts2002} Aerts, D.: Reality and Probability: Introducing a New Type of Probability Calculus. In: Aerts, D., Czachor, M., Durt, T. (eds.) Probing the Structure of Quantum Mechanics: Nonlinearity, Nonlocality, Probability and Axiomatics. pp. 205-229. World Scientific, Singapore (2002)

\end{thebibliography}
\end{document}